\documentclass[aps,prl,twocolumn,floats,superscriptaddress,showpacs]{revtex4}
\usepackage{graphicx}
\usepackage{amsmath}
\usepackage{bm}

\usepackage[dvips,usenames]{color}

\begin{document}
\bibliographystyle{myprb}
\newcommand\vf{ v _{\mathrm{  F}}}
\newcommand\kf{ k _{\mathrm{  F}}}
\newcommand\ef{ \varepsilon  _{\mathrm{  F}}}
\newcommand\nf{ n  _{\mathrm{  F}}}
\newcommand\e { \operatorname{e}  }
\newcommand\dd { \operatorname{d}  }
\newcommand\IM{\operatorname{Im}}

\title{Quantum Wire Hybridized with a Single-Level Impurity}

\author{Igor V.\ Lerner}
\affiliation{School of Physics and Astronomy, University of Birmingham,
Birmingham B15 2TT, United Kingdom}

\author{Vladimir I.\ Yudson}
 \affiliation{Institute of Spectroscopy, Russian Academy of Sciences, Troitsk, Moscow region,
142190 Russia}
\author{Igor V.\ Yurkevich}
\affiliation{School of Physics and Astronomy, University of Birmingham,
Birmingham B15 2TT, United Kingdom}

\begin{abstract}
We have studied low-temperature properties of interacting electrons
in a one-dimensional quantum wire (Luttinger liquid)  side-hybridized
with a single-level impurity. The hybridization induces a
back-scattering of electrons in the wire which strongly affects its
low energy properties. Using a one-loop renormalization group
approach valid for a weak electron-electron interaction, we have
calculated a transmission coefficient through the wire,
$\mathcal{T}(\varepsilon )$, and a local density of states,
$\nu(\varepsilon )$ at low energies $\varepsilon $. In particular,
we have found that the antiresonance in $\mathcal{T}(\varepsilon )$
has a generalized Breit-Wigner shape with the effective width
$\Gamma(\varepsilon )$ which diverges at the Fermi level.
\end{abstract}

\maketitle

Low-temperature physics of one-dimensional (1D) electron systems (quantum wires or nanotubes) is
strongly affected by electron-electron interactions. Electrons in such a
system form a Luttinger liquid (LL) \cite{Haldane:81} characterized by
power-law decay of various correlation functions
\cite{GogNersTsv,Giamarchi}.
This characteristic feature   of the LL has been  established via conductance measurements   and a scanning
tunnelling microscopy (STM)
 both  in carbon nanotubes \cite{nano} and semiconductor quantum wires \cite{QW}.

 Inserting a  potential impurity or a weak link (e.g., a tunnel barrier)
into the LL results in the power-law suppression of  a local density of states (LDoS) at the impurity site \cite{KaneFis:92a} and thus to
suppression of the conductance at low temperatures $T$, $x$-ray edge
singularity, etc.\ \cite{KaneFis:92a,MatYueGlaz:93,furusaki}. If the barrier
interrupting the LL (e.g., a quantum dot coupled to two LL leads) carries a discrete localized state, its hybridization with extended states leads to a sharp resonant transmission  \cite{KaneFis:92b}
described by a generalized Breit-Wigner formula with the
energy-dependent effective  width vanishing at the Fermi level at $T=0$.

Transmission and tunnelling measurements in the presence of controlled defects   have been performed in both quantum wires \cite{QW}  and  carbon nanotubes for various defect geometries \cite{ResTunCNWT}.
In this paper we consider how the low-$T$ electron
properties of a 1D wire are affected by the hybridization with
a discrete localized level in a  geometry where an impurity
or a quantum dot (QD) carrying such a level is side-coupled to the wire.

 When such an  impurity is hybridized with a $1D$ Fermi
gas,   the resonant level broadens to
 acquire a Lorentzian shape centered at
$\varepsilon _0$  of the width $\Gamma_0=\pi\nu_0|t_0|^2$
($\nu_0$ is the  DoS in the absence of the impurity and
$t_0$ is the tunneling amplitude). {We assume that there is only one
level close to the Fermi level $\ef$, while the level
spacing, $\delta$, on the impurity is large, $\delta\gg \Gamma_0,\,T$.} For a QD we further assume
that it is in the peak of the Coulomb blockade.
Then the electron LDoS of the Fermi gas in the vicinity of the
impurity ($x=0$) acquires a resonant dip,
\begin{eqnarray}\label{nu-0}
\frac{\nu_0(\varepsilon)}{\nu_0}= \mathcal{T}_0(\varepsilon)  =
\frac{(\varepsilon - \varepsilon_0)^2} {(\varepsilon -
\varepsilon_0)^2 + \Gamma^2_0} \,,
\end{eqnarray}
where all energies are  counted from   $\ef$.
This describes an \textit{antiresonant} structure of the
transmission coefficient $\mathcal{T}_0(\varepsilon )$. In contrast
to the resonant tunneling, it is the reflection coefficient
$\mathcal{R}_0(\epsilon)=1-\mathcal{T}_0(\varepsilon )$ which has in
this case the Breit-Wigner form, reaching the perfect reflection,
$\mathcal{R}_0(\varepsilon _0)=1$, at the resonance, $\varepsilon
=\varepsilon _0$.

As usual in the LL theory, even a weak electron-electron interaction drastically
changes both the LDoS  near the impurity and the transmission coefficient. We will
show that the transmission coefficient is given by
\begin{eqnarray}\label{T-final}
{\mathcal T}(\varepsilon) = \frac{(\varepsilon - \varepsilon_0)^2}{(\varepsilon -
\varepsilon_0)^2 + \Gamma^2(\varepsilon)}
 \,,
\end{eqnarray}
while the LDoS of conduction electrons near the impurity
($x=0$), is given by
\begin{eqnarray}\label{nu-final}
\frac{\nu(\varepsilon)}{\nu_0}  =
\frac{\Gamma(\varepsilon)}{\Gamma_0}\frac{(\varepsilon - \varepsilon_0)^2}
{(\varepsilon - \varepsilon_0)^2 + \Gamma^2(\varepsilon)}\,,
\end{eqnarray}
and the {resonant level takes the shape}
\begin{eqnarray}\label{nu-d-final}
\nu_d(\varepsilon) =
\frac{1}{\pi}\frac{\Gamma(\varepsilon)}{(\varepsilon-\varepsilon_0)^2
+ \Gamma^2(\varepsilon)}
 \, .
\end{eqnarray}
The effective level width $\Gamma( \varepsilon  )$ in Eqs.~(\ref{T-final})--(\ref{nu-d-final}) diverges in the low-energy limit
$|\varepsilon|\to 0$,
\begin{eqnarray}\label{Gamma-eff}
\Gamma(\varepsilon) =
\Gamma_0[(\Gamma^2_0 + \varepsilon^2_0)^{1/2}/|\varepsilon|]^{\alpha}
 \,,
\end{eqnarray}
and saturates at $\Gamma_0$ for $\varepsilon \gtrsim \max\left( \varepsilon
_0,\,\Gamma_0 \right) $. Here the electron-electron interaction parameter
$\alpha \equiv 1/g - 1$ is assumed to be small, $\alpha\ll1$, where $g \equiv v/\vf$
is the Luttinger parameter, $v$ ($\vf$) is the renormalized (bare)
Fermi velocity.  The divergence in Eq.~(\ref{Gamma-eff}) is cut by
temperature $T$. For $T\to0$, Eqs.~(\ref{T-final})  and (\ref{nu-final}) have a
double-dip structure with $\mathcal{T}$  and $\nu$   vanishing both at
$\varepsilon = \varepsilon _0$ and at $\varepsilon=0$.

The antiresonance described by Eq.~(\ref{T-final}) differs drastically from the
{resonance in the transmission} through a barrier \cite{KaneFis:92b} where (for
a symmetric double-barrier) {$\mathcal{T}=1$ at $\varepsilon = \varepsilon _0$,
with the resonance width \textit{vanishing} at \mbox{$\varepsilon \to0$}}. On
the contrary,   the width of the antiresonance, $\Gamma\left( \varepsilon
\right)$ in Eq.~(\ref{Gamma-eff}), \textit{diverges} at $\varepsilon \to0$.
Furthermore, the intuitively expected generalization of the Fermi-gas relation
between the level width and LDoS, i.e.\ $\Gamma\left( \varepsilon  \right)=\pi
\nu(\varepsilon )|t_0|^2$, fails at low energies for the weak-interaction
limit. Just the opposite, in this limit $\Gamma (\varepsilon )\propto
\nu^{-1}(\varepsilon )$ at $\varepsilon \to0$ while the direct proportionality
$\Gamma (\varepsilon )\propto \nu (\varepsilon )$ is restored off the
resonance, at $|\varepsilon -\varepsilon _0|\gg\Gamma\left( \varepsilon
\right)$ in Eqs.~(\ref{T-final})--(\ref{nu-d-final}).

The seemingly counter-intuitive
relation  $\Gamma (\varepsilon )\propto
\nu^{-1}(\varepsilon )$ at $\varepsilon \to0$ follows from the mapping to the case of the potential impurity: the
hybridized impurity reduces to the former in the limit $\varepsilon \ll
\varepsilon _0,\, \Gamma\left( \varepsilon  \right)$. Then both the
transmission coefficient and the LDoS should vanish at the Fermi level, $\mathcal{T}(\varepsilon
)\propto \nu^2(\varepsilon )\propto \varepsilon ^{2\alpha} $ as in
\cite{KaneFis:92a,MatYueGlaz:93}.  As it is seen from
Eqs.~(\ref{T-final}) and (\ref{nu-final}), such a behavior is ensured by the divergence in Eq.~(\ref{Gamma-eff}). Accordingly, the $T$-dependence of the conductance becomes $[\varepsilon _0/\Gamma(T)]^2\propto \varepsilon _0^2\,T^{2\alpha}$.

Such a mapping does not work for the strong electron-electron interaction in
the LL.  Indeed, let us start with the
hybridized level off the resonance with the Fermi level,  $\Gamma_0\ll\varepsilon _0 $.  In this weak hybridization limit the level
effective width is  found perturbatively in $\Gamma_0/\varepsilon _0$:
\begin{align}\label{Gstrong}
    \Gamma(\varepsilon )&=\pi\nu_{\text{bulk}}(\varepsilon )|t_0|^2\propto\varepsilon^\gamma\,,&\gamma&\equiv(1-g)^2/2g
\end{align}
Here $\nu_{\text{bulk}}(\varepsilon )$ is the electron DoS in the
pure LL system. For \mbox{$\gamma\!>\!1$}, the width of the Breit-Wigner peak is  $\Gamma(\varepsilon _0)$ and the effective perturbation parameter $\Gamma(\varepsilon _0)/\varepsilon _0$
remains small and vanishes when the peak approaches the Fermi level ($\varepsilon _0\to0$). Thus Eq.~(\ref{Gstrong})  is
self-consistent   when $\gamma\!>\!1$. In this regime the side-hybridized impurity becomes irrelevant, similar to the case of the nanowire T-junction \cite{das}, and the zero-$T$ conductance through the LL should remain ideal as in the pure LL \cite{MasStone:95}.

So, in contrast to the potential impurity case, there are two distinct regimes: the hybridized impurity makes no impact on the low-energy properties of the strongly interacting LL, while `cuts in two' (at $\varepsilon  \to 0$) a 1D wire with the weak interaction which we now consider.

We start with the Hamiltonian describing a single impurity level
hybridized with the LL:
\begin{eqnarray}\label{H}
H = H_{\textrm{LL}} + t_{0}\psi^{\dag}(0)d + {\bar
t}_{0}d^{\dag}\psi(0) + \varepsilon_0d^{\dag}d
 \, .
\end{eqnarray}
Here $H_{LL}$ is the standard  Luttinger  Hamiltonian of interacting
spinless electrons  in 1D. The LL creation and annihilation  operators are split into
the superposition of those for  right-moving ($r$) and left-moving ($\ell$)  electrons,
$\psi(x)= \psi_{r}(x)\e^{i\kf x} + \psi_{\ell}(x)\e^{-i\kf x}$, while
$d^{\dag},\,d$ are the operators for the impurity level. In what follows we consider the spinless (spin-polarized) case. A spinful case can be different, as for a very large dot side-attached to the LL \cite{kaka}, but this is beyond the scope of the current Letter.

The Green  functions for
non-interacting \textit{hybridized} conduction electrons and for $d$
electrons are given by
\begin{align}
{\hat G} _0(x,x';\varepsilon)&= {\hat g} (x-x';\varepsilon) + i\vf{\hat g}
(x;\varepsilon)\,{\hat{\mathsf
T}}_0(\varepsilon)\,{\hat g} (-x';\varepsilon)\,,\notag \\
\mathcal{G}^R_0(\varepsilon)&=  [\varepsilon - \varepsilon_0 + i\Gamma_0]^{-1}
  \, .\label{G0}
\end{align}
In the Keldysh technique  \cite{RS:86} used here both $\hat G$ and
$\hat{\mathcal{G}}$ are matrices in the Keldysh space; $\hat G$  is also a
matrix  in the \mbox{$r$-$ \ell $} space; the retarded component of the Green
function of (non-hybridized) conduction electrons is given by
\begin{align*}
    {\hat
g}^R(x;\varepsilon)&=-\frac{i}{v_F}{\hat\theta}_{x}
\e^{i\frac{\varepsilon}{v_F}|x|}\,,& {\hat\theta}_{x}&=\left(
                   \begin{array}{cc}
                     \theta(x) & 0 \\
                     0 & \theta(-x) \\
                   \end{array}
                 \right),
\end{align*}
where ${\hat\theta}_x$ is a  matrix in the
$\mbox{$r$-$ \ell $}$ space, with $\theta(x)$ being the step-function.
Here the ${\mathsf T}$-matrix  is related to the scattering matrix $\mathsf{S}$
by $\mathsf{T}\equiv\mathsf{S}-\mathsf{I}$. The bare ${\hat{\mathsf T}}$-matrix is
$
{\hat{\mathsf T}}_0^R(\varepsilon)=
r_0(\varepsilon)[\hat{1}+\,\hat{\sigma}_x],\,$
where ${\hat\sigma}_x$ is the Pauli matrix and $r_0(\varepsilon)$ and $t_0(\varepsilon)$  are the bare reflection
and transmission   amplitudes,
$
r_0(\varepsilon) =t_0(\varepsilon)-1=
 {-i\Gamma_0}\left( {\varepsilon-\varepsilon_0+i\Gamma_0} \right)^{-1}\,.
 $

The interaction corrections  to $\hat{G}$ and ${\hat{\mathcal G}}$
can be expressed via the appropriate self-energies. The self-energy
of the conduction electrons has the form
\begin{eqnarray}\label{Scal}
{\hat\Sigma} (x)=i\alpha \vf\int\dd\varepsilon\left[{\hat
G}^<_0(x,x;\varepsilon)-{\hat g}_\varepsilon^<(0)\right],
\end{eqnarray}
where the $\hat{G^<}$ component of the Keldysh Green function is related to the
retarded and advanced components via
\begin{eqnarray}
{\hat G}_0^<(x,x';\varepsilon)=-\nf (\varepsilon)\left[{\hat
G}_0^R(x,x';\varepsilon)-{\hat G}_0^A(x,x';\varepsilon)\right], \notag
\end{eqnarray}
and $\nf (\varepsilon)$ is the Fermi distribution function. The
interaction parameter $\alpha$ is given  by $\alpha =
[V(0)-V(2\kf)]/(2\pi v_F)$, with $V(0)$ and $V(2\kf)$ being the
Fourier transforms of   the
electron-electron interaction potential $V(x)$ in the forward- and back-scattering
channel.
Substituting the expression for  $\hat{G}_0^<$  into
Eq.~(\ref{Scal}) we find the self-energy of the conduction electrons in the
linear in   $\alpha$ order:
\begin{eqnarray}
{\hat\Sigma} (x)=-\alpha\,{\hat\theta}_{x}\,{\hat
\sigma}_x\,\int\dd\varepsilon'\,\nf
(\varepsilon')r_0(\varepsilon')e^{i\frac{2\varepsilon'}{v_F}|x|}
+{\mathrm h.c.}\notag
\end{eqnarray}
\begin{figure}[b]
\vspace*{2mm} \includegraphics[width=.65\columnwidth,height=33mm]{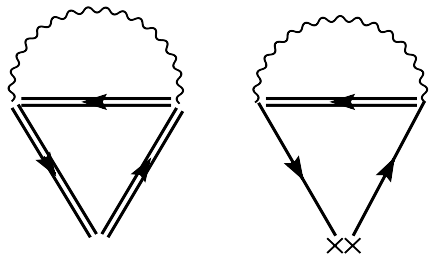}
\caption{The lowest-order Fock-type corrections:  (a) the LDoS of the conduction
electrons; (b) the $d$-electron self-energy (the appropriate Hartree corrections are not shown). Diagram (a)  with open ends  would represent
$\delta\hat{G}(x,x';\varepsilon )$. Single and double lines correspond  to the Green  functions $\hat{g}$ and $\hat{G}_0$; crosses  to the tunneling amplitudes
$t_0$;  wavy lines  to the interaction.} \label{Fock}
\end{figure}
The first-order correction to the Green  function of the conduction
electrons is then given  by
\begin{eqnarray}\label{G1}
\delta{\hat{G}^R}(x,x';\varepsilon)=\!\!\! \int\!\! d x_1
\hat{G}_0^R(x,x_1;\varepsilon){\hat\Sigma}(x_1)\hat{G}_0^R(x_1,x';\varepsilon) %\notag
\end{eqnarray}
and is graphically represented in Fig.~\ref{Fock}a. Now the
correction to the ${\mathcal T}(\varepsilon)$ is found from the
asymptotic expressions ${\hat G}^R(x,x'\!\to\!\pm\infty)$ which are
related to the $\mathsf T$-matrix in the same way as $\hat{G}_0$ to
$\mathsf T_0$, Eq.~(\ref{G0}):
\begin{eqnarray}\label{delT}
\delta{\hat{\mathsf
T}}(\varepsilon)=-\frac{\alpha}{2}\left[{\hat{\mathsf
S}}_0(\varepsilon){\hat\sigma}_x{\hat{\mathsf
S}}_0(\varepsilon)\,{\bar {\cal
L}}_0(\varepsilon)-{\hat\sigma}_x{\cal L}_0(\varepsilon)\right]
\end{eqnarray}
where
\begin{eqnarray} {\cal L}_0(\varepsilon) =\!\!\int\limits_{-\infty}^{+\infty}\frac{\nf
(\varepsilon')\, \dd\varepsilon'}{\varepsilon-\varepsilon'}r_0(\varepsilon')
\simeq r_0(\varepsilon)\,\ln\frac{\varepsilon_0-i\Gamma_0}{\max(\varepsilon,
T)}. \label{L-0}
\end{eqnarray}
 From Eq.~(\ref{delT}) we find the corrections   to $r_0$ and $t_0$:
\begin{align}\label{Corr-t-r}
\delta t= -\alpha\,r_0\,t_0{\bar{\cal L}}_0\,,\quad \delta r =
-\frac{\alpha}{2}\,\left[\left(r^2_0+t^2_0\right){\bar{\cal
L}}_0-{\cal L}_0
 \right]\, .
\end{align}
We focus further considerations on the strong-hybridization case,  $\varepsilon_0\ll\Gamma_0$, as in the opposite case an impact of the
resonant impurity on the LDoS takes place only in a very narrow region $\varepsilon \lesssim
\Gamma_0 (\Gamma_0/\varepsilon _0)^{1/\alpha}$.

The  divergence at $(\varepsilon,T)\to0$ in  Eq.~(\ref{L-0}) necessitates a
resummation of the main logarithmic corrections in all orders which is
performed in the weak-interaction limit within the standard renormalization
group (RG) approach. The appropriate scheme was elaborated in
Ref.~\cite{MatYueGlaz:93} and used also  for dealing with a resonant scatterer
inside the LL \cite{KaneFis:92b}. A crucial difference
comes  in the present case from the   {\it anti}-resonant character of the scatterer which means that the
reflection amplitude, $r_0(\varepsilon)$, vanishes at the energies higher then
$\Gamma_0$. Hence, there are no two different regimes present in the
resonant case \cite{KaneFis:92b} (the low-energy one with $\varepsilon \lesssim \Gamma_0$ and the
high-energy one, with $\varepsilon $ between $\Gamma_0$ and the bandwidth) but
only the low-energy one where generic integro-differential RG equations become
ordinary differential equations.

Following the same line of reasoning as in \cite{MatYueGlaz:93,KaneFis:92b} we
integrate out the energy window $[\Gamma_0/\lambda,\Gamma_0]$  with $\lambda$
being a running cut-off. The RG flows start at $\lambda=1$,  with the initial
conditions $\mathcal{T}=\mathcal{T}_0,\,\,\Sigma_d^R=-i\Gamma_0$, and stop at
$\lambda = \Gamma_0/|\varepsilon|$.  The corresponding RG equations for the
scattering amplitudes take the standard form:
\begin{align}
\frac{\dd\ln{t(\varepsilon,\lambda)}}{\dd \ln{\lambda} } &=
-\alpha\,\mathcal{R}\, , &
 \frac{\dd\ln{r(\varepsilon,\lambda)}}{\dd \ln{\lambda} } &=
\alpha\,\mathcal{T}\,.\label{RG-rt}
\end{align}
The solution to Eq.~(\ref{RG-rt}) gives the transmission coefficient ${\mathcal
T}(\varepsilon)=|t(\varepsilon)|^2$, Eq.~(\ref{T-final}). Note that if one
changed the initial conditions  in  Eq.~(\ref{RG-rt}) from the
antiresonance to the resonance (when it is $\mathcal{T}_0$ rather
than $\mathcal{R}_0$  has the Breit-Wigner form), then the solution to
these equations would coincide with that obtained  for the symmetric
double-barrier \cite{KaneFis:92b}, but with the high-energy cutoff at
$\Gamma_0$.

The correction to the self-energy $\Sigma_{d0}^R$ of the impurity
Green function is represented in Fig.~\ref{Fock}b and has the form:
\begin{eqnarray}\label{Corr-Sigma}
\delta{\Sigma_d^R}(\varepsilon)=-\alpha\, \Sigma_{d0}^R \,{\bar{\cal
L}}(\varepsilon) \,
\end{eqnarray}
Then in a similar way one derives from the perturbative expression (\ref{Corr-Sigma}) the RG equation
for $\Sigma_d$:
\begin{align}\label{RG-Sigma}
\frac{\dd\ln{\Sigma_d^R(\varepsilon,\lambda)}}{\dd \ln{\lambda} } =
\alpha\,
\end{align}
Such a ``poor man" RG approach can  be justified through the use of
a functional bosonization technique in the form suggested in
\cite{GYL:04}.  In contrast to the standard operator bosonization,
it is suitable for making the weak interaction expansion.  In this
framework, one can integrate out the conduction electrons degrees of
freedom for $\alpha\ll1$. The RG analysis of the effective
d-electron  action confirms the validity of the RG equation
(\ref{RG-Sigma}). Solving this equation results in the following
energy dependence of the $d$ electron level width:
\begin{align}\label{Sigma-fin}
\Sigma^R_d(\varepsilon) =
-i\Gamma_0(\Gamma_0/|\varepsilon|)^{\alpha} \equiv
-i\Gamma(\varepsilon) \, ,
\end{align}
where $\Gamma(\varepsilon)$ is given by Eq.(\ref{Gamma-eff}). The
impurity DoS around the resonant level is then $\nu_d(\varepsilon)\! =\!
-(1/\pi)\mathrm{Im}[\varepsilon - \varepsilon_0 - \Sigma_d^R]^{-1}$, which results in Eq.~(\ref{nu-d-final}).

To find the LDoS of the LL electrons in the vicinity of the impurity we use the exact relationship between the
impurity Green  function, $\mathcal{G}^R(\varepsilon) =
[\varepsilon-\varepsilon_0 -\Sigma_d^R(\varepsilon)]^{-1}$, and the
local Green  function, ${G}^R(\varepsilon)$, of the LL electrons,
\begin{eqnarray}\label{Dyson}
\mathcal{ {G}}^R(\varepsilon) =
\mathcal{ {G}}^R_{00}(\varepsilon) +
\mathcal{ {G}}^R_{00}(\varepsilon)|t|^2 {G}^R(\varepsilon)
\mathcal{ {G}}^R_{00}(\varepsilon)
 \, ,
\end{eqnarray}
where $ {G}^R(\varepsilon) =
\sum_{\eta,\eta'} {G}^R_{\eta\eta'}(0,0;\varepsilon)$ with
$\eta,\eta'=r,\ell$ and   $\mathcal{G}^R_{00}(\varepsilon) = (\varepsilon -
\varepsilon_0 + i0)^{-1}$. Using this identity one can prove the exact
relationship between the LDoS of the LL electrons at $x=0$,
$\nu(\varepsilon) = -(1/\pi)\mathrm{Im}G^R(\varepsilon)$, and the
impurity DoS $\nu_d(\varepsilon)$:
\begin{eqnarray}\label{id-nu}
(\varepsilon-\varepsilon_0)^2\,\nu_d(\varepsilon)=|t|^2\,\nu(\varepsilon) \, .
\end{eqnarray}
Together with Eq.~(\ref{nu-d-final}) this identity leads to the
expression (\ref{nu-final}) for the LDoS of the conduction electrons
at $x=0$.  It shows a double-dip at $\varepsilon =\varepsilon _0$ and $
\varepsilon =0$ in $\nu\left( \varepsilon  \right)$ which can  be observed for
 $T\lesssim \varepsilon _0, \, \Gamma_0\operatorname e^{-1/\alpha}$.

The RG equations (\ref{RG-rt}),\,(\ref{RG-Sigma}) are valid
only in the region $|\varepsilon|,\varepsilon_0 \ll \Gamma_0$ (where
the expansion in $\alpha$ contains logarithmic divergences) while
outside this region the interaction corrections are
non-singular. Therefore, Eqs.~(\ref{nu-final}) and
(\ref{nu-d-final}) could be considered as possible interpolation formulae
which are asymptotically exact both for low energies (the RG regime) and for high energies (the Fermi gas limit).

Now we briefly address the problem of the resonant-level occupation, $n(\varepsilon_0)$. This problem has first been studied for the resonant level {side}-hybridized   with a
\textit{chiral} LL  \cite{FurMatv:02}, and later \cite{Sade} for the level  in a QD connecting two {\it non-chiral} LL's. The present case of the level side-hybridized   with a
\textit{non-chiral} LL differs from that in \cite{FurMatv:02} by the back-scattering induced change in the LDoS, Eq.~(\ref{nu-final}). However,  in the weak-interaction limit
    $\nu_d(\varepsilon )$, Eq.~(\ref{nu-d-final})
substantially deviates from the Lorentzian only in a small region,
$\varepsilon \lesssim \Gamma_0\operatorname e^{-1/\alpha}$. Thus $n(\varepsilon_0)$ differs  only  by a small correction from that for the Fermi gas, $n(\varepsilon_0)=1/2- (1/\pi)\arctan(\varepsilon _0/\Gamma_0)$. In the strong interaction limit,  $\Gamma(\varepsilon )$ in $\nu_d(\varepsilon )$ is given by Eq.~(\ref{Gstrong}) for $\gamma>1$.  Then $\nu_d(\varepsilon )$ is sharply peaked around $\varepsilon _0$  with the width $ \Gamma(\varepsilon _0)\ll\varepsilon _0$ at $\varepsilon_0 \to0$.  This leads to the expression for $n(\varepsilon _0)$ like for the Fermi gas above, but with $\Gamma_0\mapsto\Gamma(\varepsilon _0)$. Since $|\varepsilon _0|/\Gamma(\varepsilon _0)\to\infty$ at $\varepsilon _0\to0$, the occupation $n(\varepsilon _0)$ has the Fermi jump  at $\varepsilon _0=0$.
This is  in agreement with the explanation after Eq.~(\ref{Gstrong}) that the $d$-level remains off the resonance with the Fermi level in the wire  at $\varepsilon _0\to0$.

Low-$T$ resonant transport in the geometry similar to that considered  in this Letter (a QD side-coupled to a single-channel quantum wire) has already been experimentally investigated \cite{Side}. To detect Luttinger features described here further transport and STM measurements are required similar to those performed in \cite{QW}.

To conclude, we have studied low-$T$ properties of the non-chiral
LL hybridized with a side-attached impurity carrying a single resonant level.  The results for the transmission through the LL and the electron LDoS in the LL and on the impurity are given by Eqs.~(\ref{T-final})--(\ref{Gstrong}).    The antiresonance in the transmission coefficient has a double-dip described by the generalized Breit-Wigner  formula with the effective width $\Gamma(\varepsilon )$ that diverges at the Fermi level  ($\varepsilon\! \to\!0$) in  the weak interaction limit, Eq.~(\ref{Gamma-eff}), while vanishes for the strong electron-electron interaction, Eq.~(\ref{Gstrong}). This is in a striking contrast to the known results for the resonant transmission through a (symmetric) double-barrier, where the width of the peak always vanishes at the Fermi level \cite{KaneFis:92b}. Furthermore, in the vicinity of the Fermi level, the intuitively expected relation $\Gamma(\varepsilon )\propto \nu(\varepsilon )$  holds only for the strong interaction, with $\nu(\varepsilon )$ being the intrinsic DoS in the LL. For the weak interaction the inverse proportionality takes place, $\Gamma(\varepsilon )\propto \nu^{-1}(\varepsilon )$, with $\nu(\varepsilon )$ being LDoS modified by the impurity, Eq.~(\ref{nu-final}) and (\ref{Gamma-eff}). Such a behavior of $\Gamma(\varepsilon )$ indicates the existence of a quantum phase transition in the interaction strength: the side-attached impurity does not affect low-$T$ properties of the LL with the strong electron-electron interaction, while results in the zero conductance and LDoS at the Fermi level for the weak interaction.					

\begin{acknowledgments}
We thank B.~L.~Altshuler, I.~L.~Aleiner, S.~Carr, L.~I.~Glazman,
A.~Kamenev and M.~Khodas for helpful discussions.  We acknowledge  support by EPSRC  (grant EP/D031109), the Royal Society  and
 the RFBR (grant 06-02-16744, V.Yu.) and 
hospitality of the Abdus Salam ICTP where part of the work was done.
\end{acknowledgments}

%\bibliography{my,ag01}\end{document}

\end{document}